\documentclass[letterpaper,journal]{IEEEtran}
\usepackage{amsmath,amsfonts}
\usepackage{algorithmic}
\usepackage{algorithm}
\usepackage{array}
\usepackage[caption=false,font=normalsize,labelfont=sf,textfont=sf]{subfig}
\usepackage{textcomp}
\usepackage{stfloats}
\usepackage[hyphens]{url}
\usepackage[hidelinks]{hyperref}
\usepackage{verbatim}
\usepackage{graphicx}
\usepackage{cite}

\usepackage[dvipsnames]{xcolor}
 
\usepackage{glossaries}
\usepackage{booktabs}
\usepackage{multirow}
\usepackage{tabularx}
\usepackage{makecell}
\newcolumntype{Y}{>{\centering\arraybackslash}X} 

\newacronym{3GPP}{3GPP}{3rd Generation Partnership Project}
\newacronym{5GC}{5GC}{5G Core}
\newacronym{5GAA}{5GAA}{5G Automotive Association}
\newacronym{5GCAR}{5GCAR}{5G Communication Automotive Research and innovation}
\newacronym{AI}{AI}{Artificial Intelligence}
\newacronym{AMF}{AMF}{Access and Mobility Management Function}
\newacronym{AUSF}{AUSF}{Authentication Server Function}
\newacronym{CBR}{CBR}{Constant Bit Rate}
\newacronym{CN}{CN}{Core Network}
\newacronym{CQI}{CQI}{Channel Quality Indicator}
\newacronym{CU}{CU}{Central Unit}
\newacronym{C-V2X}{C-V2X}{Cellular V2X}
\newacronym{DL}{DL}{Downlink}
\newacronym{DRX}{DRX}{Discontinuous Reception}
\newacronym{DRL}{DRL}{Deep Reinforcement Learning}
\newacronym{DSRC}{DSRC}{Dedicated Short Range Communications}
\newacronym{DU}{DU}{Distributed Unit}
\newacronym{E2E}{E2E}{end-to-end}
\newacronym{eMBB}{eMBB}{Enhanced Mobile Broadband}
\newacronym{FDD}{FDD}{Frequency Division Duplex}
\newacronym{KPI}{KPI}{Key Performance Indicator}
\newacronym{KQI}{KQI}{Key Quality Indicator}
\newacronym{LoA}{LoA}{Level of Automation}
\newacronym{LTE}{LTE}{Long-Term Evolution}
\newacronym{MAC}{MAC}{Medium Access Control}
\newacronym{MEC}{MEC}{Multi-access Edge Computing}
\newacronym{ML}{ML}{Machine Learning}
\newacronym{mMTC}{mMTC}{massive Machine Type Communications}
\newacronym{Near-RT}{Near-RT}{near-Real Time}
\newacronym{NFV}{NFV}{Network Function Virtualization}
\newacronym{NSSF}{NSSF}{Network Slice Selection Function}
\newacronym{Non-RT}{Non-RT}{Non-Real Time}
\newacronym{OAM}{OAM}{Operation, Administration and Management}
\newacronym{OSS}{OSS}{Operations Support System}
\newacronym{PDCP}{PDCP}{Packet Data Convergence Protocol}
\newacronym{PDU}{PDU}{Protocol Data Unit}
\newacronym{PM}{PM}{Performance Management}
\newacronym{PoC}{PoC}{Proof of Concept}
\newacronym{PPO}{PPO}{Proximal Policy Optimization}
\newacronym{PRB}{PRB}{Physical Resource Block}
\newacronym{QoS}{QoS}{Quality of Service}
\newacronym{RAN}{RAN}{Radio Access Network}
\newacronym{rApp}{rApp}{RAN Application}
\newacronym{RL}{RL}{Reinforcement Learning}
\newacronym{RLC}{RLC}{Radio Link Control}
\newacronym{RSU}{RSU}{Road Side Unit}
\newacronym{RT}{RT}{Real Time}
\newacronym{SA}{SA}{Standalone}
\newacronym{SBA}{SBA}{Service Based Architecture}
\newacronym{SCS}{SCS}{Subcarrier Spacing}
\newacronym{SDN}{SDN}{Software Defined Networking}
\newacronym{SLA}{SLA}{Service Level Agreement}
\newacronym{SMF}{SMF}{Session Management Function}
\newacronym{SMO}{SMO}{Service Management and Orchestration}
\newacronym{SON}{SON}{Self-Organizing Network}
\newacronym{SLO}{SLO}{Service Level Objectives}
\newacronym{SST}{SST}{Slice Service Type}
\newacronym{TDD}{TDD}{Time Division Duplex}
\newacronym{UDM}{UDM}{Unified Data Management}
\newacronym{UDP}{UDP}{User Datagram Protocol}
\newacronym{UL}{UL}{Uplink}
\newacronym{UPF}{UPF}{User Plane Function}
\newacronym{URLLC}{URLLC}{Ultra-Reliable Low Latency Communications}
\newacronym{V2I}{V2I}{Vehicle-to-Infrastructure}
\newacronym{V2N}{V2N}{Vehicle-to-Network}
\newacronym{V2N2V}{V2N2V}{Vehicle-to-Network-to-Vehicle}
\newacronym{V2P}{V2P}{Vehicle-to-Pedestrian}
\newacronym{V2V}{V2V}{Vehicle-to-Vehicle}
\newacronym{V2X}{V2X}{Vehicle-to-Everything}
\newacronym{VBR}{VBR}{Variable Bit Rate}
\newacronym{VRU}{VRU}{Vulnerable Road User}
\newacronym{xApp}{xApp}{eXtended Application}

\glsdisablehyper
\begin{document}

\title{DRL-driven RAN Slicing Management: A V2X-oriented Approach In Multi-service Scenarios}

\author{Daniel E. Garcia-Fernandez, Pablo Vera-Soto, Sergio Fortes,~\IEEEmembership{Senior Member,~IEEE}, \\ M. Martinez, I. de-la-Bandera, M. L. Luque, A. Mendo, J. Ramiro, Raquel Barco
\thanks{This work has been submitted to the IEEE for possible publication. Copyright may be transferred without notice, after which this version    may no longer be accessible. This work has been partially funded by Ministerio para la Transformaci\'on Digital y de la Función Pública and European Union - NextGenerationEU within the framework ``Recuperaci\'on, Transformaci\'on y Resiliencia'' under the project 5GVEC.}
\thanks{D. E. Garcia-Fernandez, Pablo Vera-Soto, Sergio Fortes, M. Martínez, I. de-la-Bandera, and Raquel Barco are with the Telecommunications Research Institute
(TELMA), E.T.S.I de Telecomunicación, University of Málaga, 29010 Málaga,
Spain (e-mail: danielgarciaf@uma.es; pablovera@uma.es; sergio.fortes@uma.es; mcruces@uma.es; ibanderac@uma.es; rbarco@uma.es).}
\thanks{M. L. Luque, A. Mendo, and J. Ramiro are with Ericsson, C Severo Ochoa 55, 29590 Málaga, Spain (e-mail: maria.laura.luque@ericsson.com, adriano.mendo@ericsson.com, juan.ramiro@ericsson.com).
}
}


\makeatletter
\def\ps@myIEEEstyle{%
  \def\@oddhead{%
    \scriptsize
    This work has been submitted to the IEEE for possible publication. Copyright may be transferred without notice, after which this version    may no longer be accessible.
    \hfill\thepage
}%
  \def\@evenhead{%
    \scriptsize
    This work has been submitted to the IEEE for possible publication. Copyright may be transferred without notice, after which this version    may no longer be accessible.
    \hfill\thepage
}%
  \def\@oddfoot{}%
  \def\@evenfoot{}%
}
\makeatother

\maketitle

\pagestyle{myIEEEstyle}
\thispagestyle{myIEEEstyle}

\begin{abstract}
    The integration of \gls{V2X} communications is driving a profound transformation in vehicular connectivity, expected to significantly enhance traffic efficiency and safety. However, the stringent requirements of V2X services, particularly ultra-low latency and high reliability, present significant technical challenges. 5G's Network Slicing emerges as a key enabler by providing tailored virtual networks that ensure isolation and adaptability for heterogeneous services. This work proposes an intelligent Radio Access Network (RAN) slicing management framework specifically designed for scenarios where safety-critical V2X and high-capacity \gls{eMBB} slices coexist. In such complex environments, harmonizing conflicting traffic requirements demands continuous, data-driven optimization. To achieve this, the proposed framework leverages an advanced \gls{DRL} approach which dynamically optimizes resource allocation in real time. The framework is empirically validated on a real 5G Standalone (SA) network, where experimental results demonstrate that the DRL-driven approach successfully balances both objectives, outperforming traditional static and proportional allocation strategies by minimizing SLA violations while ensuring high resource utilization for eMBB slices.
    \end{abstract}

    
\section{Introduction}
\IEEEPARstart{T}{he} automotive industry is undergoing a major paradigm shift driven by the advent of connected vehicles. Leveraging Vehicle-to-Everything (V2X) communication technologies, this transition promises to drastically improve road safety and traffic efficiency. However, alongside ambitious goals such as autonomous and remote driving, new challenges emerge, particularly associated with the demanding nature of these communications, which are conditioned by high mobility and a trend toward safety-critical services. 

\begin{figure*}[t]
    \centering
    \includegraphics[width=0.98\textwidth]{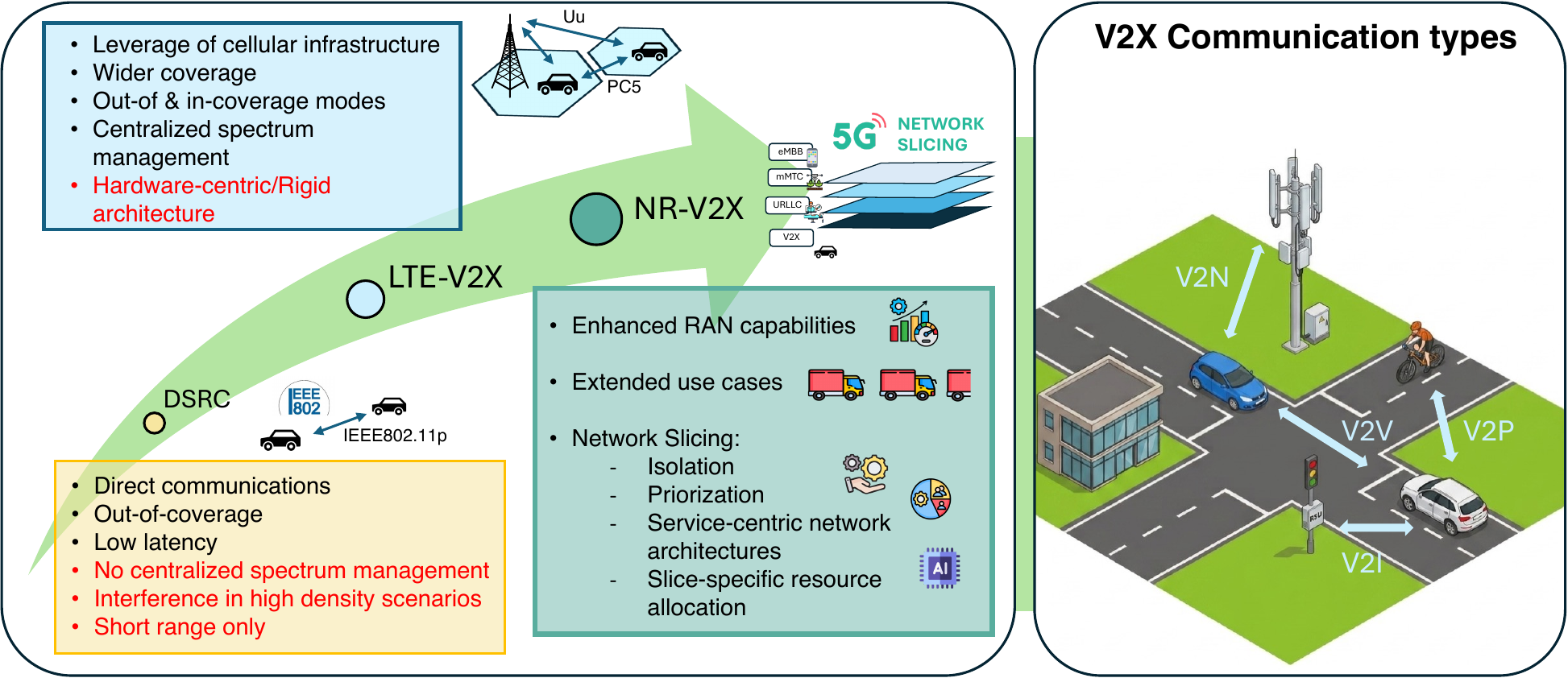}
    \caption{Left: Evolution of V2X communication technologies. Right: Communication types in Cellular V2X.}
    \label{fig:v2x-evolution}
\end{figure*}

While initially based on short-range direct communications standardized in IEEE 802.11p, commonly known as Dedicated Short Range Communications (DSRC), this early technology faced significant limitations regarding communication range, scalability, and interference in high-density scenarios \cite{DSRC_density}, limiting the deployment of advanced use cases such as autonomous driving. These limitations fostered a shift towards cellular networks. The evolution into Cellular V2X (C-V2X), illustrated in Fig. \ref{fig:v2x-evolution}, enables the four fundamental communication paradigms: Vehicle-to-Vehicle (V2V), Vehicle-to-Infrastructure (V2I), Vehicle-to-Pedestrian (V2P), and Vehicle-to-Network (V2N). 

In this way, the integration of cellular technology offers significant advantages, including centralized radio resource management, wider coverage, extended environmental awareness, and the enablement of remote server applications. However, a major challenge in integrating V2X into mobile networks is adapting to heterogeneous services with diverse requirements, especially considering the \textit{one-size-fits-all} architecture of \gls{LTE} technology. To address this, 5G introduces Network Slicing as a native feature, enabling the creation of multiple virtual networks over a single physical infrastructure, each tailored to specific service needs.

Slicing enables service-centric architectures where multiple tenants can use common resources with strict isolation, reducing Capital Expenditures (CAPEX) and Operating Expenses (OPEX) \cite{CAPEX-OPEX}. Despite its potential, implementing Network Slicing introduces severe complexities in terms of dynamic resource management and optimization, particularly in the Radio Access Network (RAN) where the physical resources are highly contended. Given the strict optimization demands of V2X requirements, intelligent slicing management systems are essential. Traditional iterative algorithms struggle to adapt to the highly dynamic radio environment \cite{Meta-TraditionalvsDRL}. Consequently, Machine Learning (ML), and specifically Deep Reinforcement Learning (DRL), presents a promising solution due to its ability to handle large amounts of data and learn optimal policies through interaction in complex environments \cite{DRLforRM}.

Furthermore, as highlighted in recent surveys \cite{Ebrahimi_E2E_2024}, while Network Slicing has been extensively studied in the Core Network (CN) segment where resources are primarily computational, the true bottleneck for end-to-end Service Level Agreement (SLA) compliance resides in the RAN due to spectrum scarcity. In fact, the lack of commercial solutions supporting near-real-time, multi-domain slicing remains a major barrier for operators to offer full 5G slicing services.

While the coexistence of safety-critical V2X and high-bandwidth enhanced Mobile Broadband (eMBB) services has been increasingly explored in recent literature, existing approaches present significant limitations. Several studies either lack the explicit enforcement of strict SLAs \cite{NO-SLA-NO-RL}, or rely on traditional allocation methods instead of intelligent, learning-based approaches \cite{NO-RL}. However, among the works that do incorporate intelligent mechanisms to ensure SLA compliance under dynamic conditions, practical deployments remain severely limited, as highlighted in \cite{NoEmpirical}. Consequently, the majority of these intelligent solutions are validated solely through simulation \cite{PREPRINT-SIM} \cite{SIMULATIONS}, lacking empirical evidence of their feasibility and performance on real-world 5G infrastructure, where physical layer constraints and operational delays heavily dictate the system's effectiveness.

To bridge these gaps, this work proposes an intelligent slice management framework powered by a Proximal Policy Optimization (PPO) DRL algorithm for dynamic resource allocation at the cell level. The novel design enables the coexistence of multiple heterogeneous services by jointly optimizing strict SLA fulfillment and spectral efficiency. Specifically, the design targets complex scenarios where safety-critical V2X slices with stringent latency Service Level Agreements (SLAs) must operate alongside capacity-hungry eMBB slices. The proposed DRL agent resolves this conflict by treating V2X SLA compliance as the primary optimization objective, while simultaneously maximizing the overall throughput for eMBB traffic. Furthermore, to demonstrate its practical viability, the framework is thoroughly evaluated through a closed-loop deployment on a real-world 5G Standalone (SA) network based on commercial equipment, proving improved performance over traditional allocation strategies.
    
\section{RL for slicing in V2X scenarios}

Network Slicing, fundamentally enabled by Network Function Virtualization (NFV) and Software-Defined Networking (SDN) \cite{NFV-SDN}, can support critical verticals like V2X by provisioning isolated virtual networks over a shared infrastructure. However, orchestrating diverse resources across multiple network segments and management levels significantly elevates complexity, which is increasingly being addressed through DRL approaches for dynamic resource allocation.

\subsection{Resource Management for Slicing}
\label{sec:problem-statement}
The lifecycle of a network slice typically comprises four phases: preparation, commissioning, operation, and decommissioning \cite{SliceLifecycle}. While commissioning involves instantiating the slice and allocating the initial resources required for its functioning, the operation phase encompasses its active exploitation. During this last phase, control spans across the three main network segments: Core, Transport, and Radio Access Network (RAN). Each one of these segments provides diverse resource types across them to satisfy specific slice requirements, as shown in Fig. \ref{fig:resource-types}.

While some resource configurations (e.g., core function placement) are relatively static, allowing centralized allocations over broader temporal and geographical scopes, radio resources are highly dynamic. For this reason, a combination of centralized and local management is widely proposed \cite{LocalCentralized-1}, operating at different timescales, namely near Real-Time and non Real-Time, to balance general orchestration with rapid localized adaptations.

\subsection{Reinforcement Learning for Dynamic Allocation}

The highly dynamic nature of the radio environment, exacerbated by the high mobility of V2X services, makes traditional optimization methods impractical for real-time RAN slicing management. 
Consequently, DRL has emerged as a powerful alternative capable of handling the high-dimensional state-action spaces typical of complex wireless networks.

Among DRL algorithms, PPO, first introduced by OpenAI in \cite{PPO}, stands out for its stability, ease of implementation, and high sample efficiency. This last characteristic is especially important because data is often the bottleneck in AI-driven network management, given the scarcity of datasets provided by operators, alongside the practical risks of training Reinforcement Learning (RL) agents online in a live network. PPO utilizes an Actor-Critic architecture where the Actor network determines the policy (the probability distribution of actions) and the Critic network evaluates the value of the current state. Unlike simpler policy gradient methods, PPO incorporates a clipping mechanism in its objective function to discourage excessively large policy updates. This inherent stability is especially well-suited for network management scenarios, where erratic policy shifts could temporarily collapse the service quality of active slices. 

\section{V2X-oriented Intelligent RAN Slicing Framework}

When it comes to slicing management, Over-The-Top (OTT) or purely centralized operator level solutions offer higher signaling overheads and centralized bottlenecks, while vendor-level approaches can provide deeper integration with the underlying network infrastructure. Furthermore, while both centralized and local management approaches are important in 5G networks, a local, RAN-centric approach is particularly suitable for radio resource allocation due to the rapid dynamism of the wireless channel.

Therefore, this work proposes a DRL-driven, cell-level intelligent resource management framework structured into three independent layers, namely Business, Network, and Management, enabling a robust closed-loop control over the RAN spectrum.

\begin{figure}[t]
    \centering
    \includegraphics[width=0.98\columnwidth]{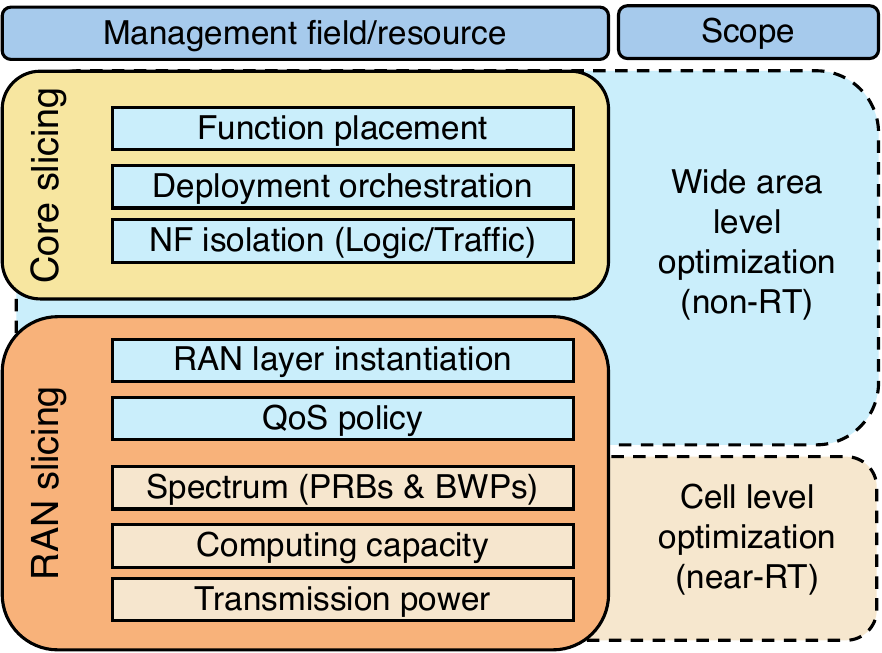}
    \caption{Configurable network resource types for Core and RAN segments.}
    \label{fig:resource-types}
\end{figure}

\begin{figure*}[t]
    \centering
    \includegraphics[width=0.98\textwidth]{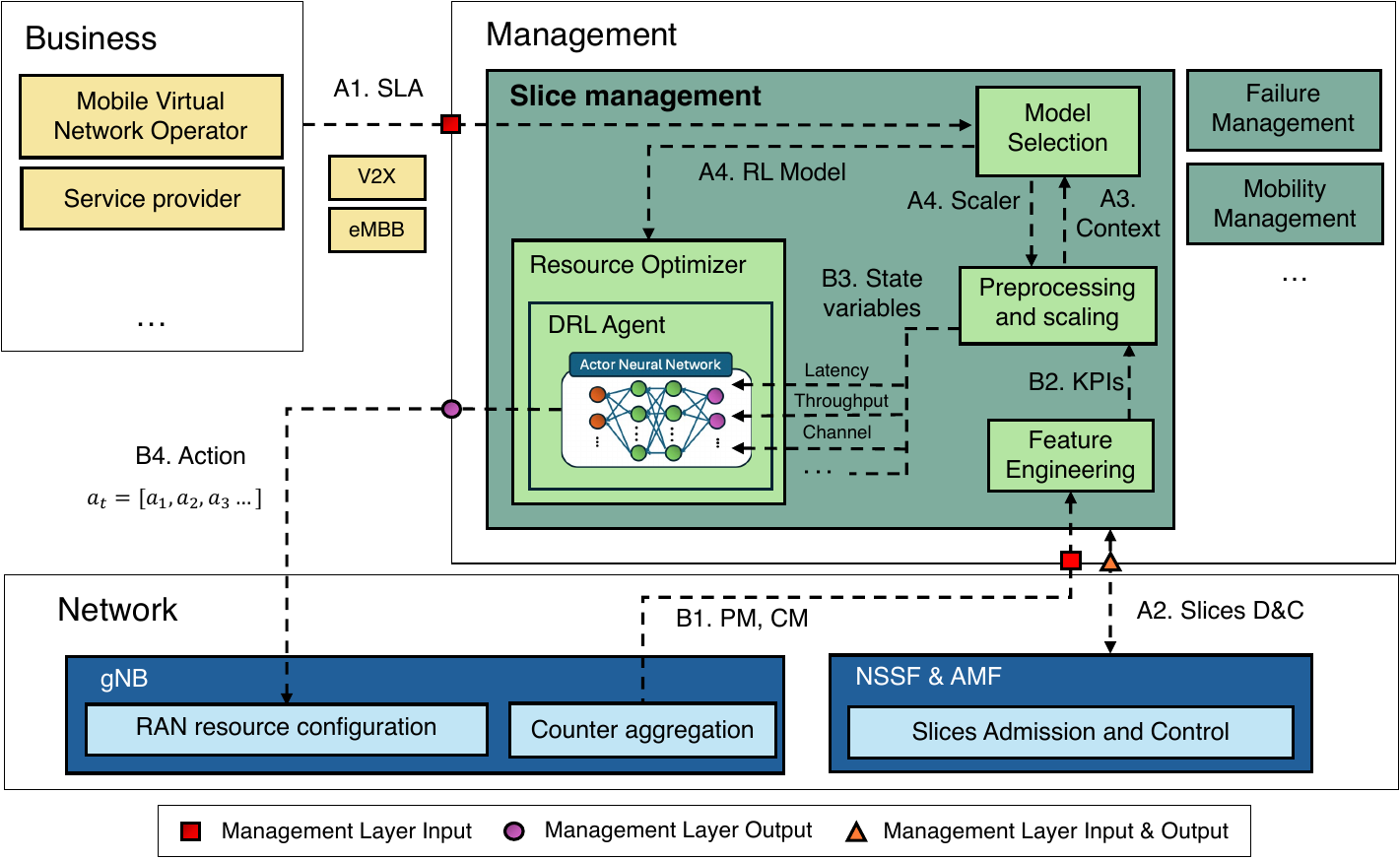}
    \caption{Architecture of the Intelligent RAN Slicing Framework, highlighting the Business, Network, and Management layers alongside the RL Agent loop.}
    \label{fig:framework}
\end{figure*}

\subsection{Architecture Layers}

The proposed architecture adopts a local, RAN-centric approach structured in three independent layers, as illustrated in Fig. \ref{fig:framework}. Parallel to the slicing management lifecycle, the framework contemplates an initial deployment and configuration phase, where network slices are instantiated and a specific DRL model is selected depending on the number of services and their requirements. This sequence is represented by steps A1-A4. In a second phase, associated with management and operation, the closed resource management loop represented by steps B1-B4 is executed. Here, resource partitions are dynamically configured on the network based on the state variables observed by the agent. 

\subsubsection{Business Layer}
In this layer, multiple tenants, such as Mobile Virtual Network Operators (MVNOs) or vertical industries (e.g., automotive), request network resources. A critical function of this level is the accurate extraction of Service Level Agreements (SLAs) prior to slice deployment, ensuring the network can support the requested services. Moreover, challenges regarding resource allocation appear in congestion scenarios, where simultaneous fulfillment of all demands becomes unfeasible. It is at this point where the Business Layer needs to establish strict priority policies among tenants. 

For the specific scenario addressed in this work, V2X slices are bound to strict latency thresholds, whereas eMBB throughput maximization is considered as a secondary objective. Consequently, the primary goal of the system is to provide an accurate resource allocation that targets V2X SLAs without over-provisioning, later ensuring eMBB maximization once V2X requirements are met.  

\subsubsection{Network Layer}

This layer encompasses the underlying 5G infrastructure, integrating both Core and RAN Network Elements (NEs). Initially, slice instantiation takes place, which involves the Deployment and Configuration (D\&C) of end-to-end (E2E) resources. This includes essential Core functions such as the Access and Mobility Management Function (AMF) and the Network Slice Selection Function (NSSF), essential for user admission control and session routing. During the operational phase, however, ensuring dynamic performance requires precise control over the radio interface. Among the various configurable resources illustrated in Fig. \ref{fig:resource-types}, this work focuses on the partition of Physical Resource Blocks (PRBs) within the RAN.

To manage spectrum allocation efficiently, the framework leverages a \textit{soft-slicing} paradigm based on PRB quotas. Unlike \textit{hard-slicing}, which imposes strict isolation boundaries between slices, soft-slicing permits spectral resources to be dynamically shared. PRB quotas are defined by a maximum and a minimum threshold. The maximum quota dictates the upper limit of resources a slice can consume; in this work, it is set to 100\% for all slices, ensuring that no available resources are artificially wasted and any spare capacity can be reused. Conversely, the minimum quota acts as a guaranteed lower bound, and dynamically adjusting these minimum quotas is what effectively dictates the prioritization policy among the active slices.

The primary objective of this dynamic partitioning is to enforce the desired prioritization policy (defined at the Business Layer) at the radio level, preventing the base station's default scheduling algorithms from unpredictably redistributing spare resources. Finally, the Network Layer is responsible for continuous telemetry collection. Given the disaggregated 5G RAN architecture, this involves gathering Configuration Management (CM) parameters and Performance Management (PM) counters across multiple protocol layers, capturing physical resource usage and channel quality at the Distributed Unit (DU), alongside flow-oriented statistics at the Central Unit (CU).

\subsubsection{Management Layer}
At the core of the framework, this layer operates from a vendor's perspective, with direct access to network metrics and slice configuration mechanisms. Its primary objective is the dynamic optimization of radio resources to adapt to the observed network state. Following the Deployment and Operation phases, the initial step involves selecting the most appropriate RL model based on the cell context, the number of active services, and their specific requirements. This cell context is determined by several factors, including current cell load, interference levels, active slices, and targeted SLAs.

Internally, the Management Layer is divided into four key operational blocks:
\begin{itemize}
    \item \textbf{Feature Engineering:} Raw CM parameters and PM counters collected from the Network Layer are processed and mathematically transformed into meaningful Key Performance Indicators (KPIs). This processing involves filtering erratic values, as well as obtaining high-level metrics from raw data formats; for example, estimating the average \textit{Protocol Data Unit} (PDU) delay at the Radio Link Control (RLC) layer by aggregating cumulative delays and the total number of transmitted PDUs. 
    \item \textbf{Model Selection:} Contextual parameters, such as cell load and neighbor interference, are retrieved from RAN telemetry provided by the Network Layer, whereas the number of active services and their requirements are provided by the Business Layer. Utilizing this information, the Model Selection block chooses a pre-trained RL model from a catalog, along with the corresponding input scaler used during model training.
    \item \textbf{Preprocessing and Scaling:} The generated KPIs undergo filtering to eliminate outliers and are subsequently normalized. This scaling process utilizes the statistical parameters (e.g., mean and variance) and outlier removal criteria derived during the training phase of the selected RL model.
    \item \textbf{Resource Optimizer:} Acting as the decision-making engine, this block loads and executes the selected RL model. It drives a continuous closed-loop interaction with the environment by evaluating state observations from the Network Layer and issuing the optimal PRB quota adjustments to meet the optimization objectives.
\end{itemize}

\subsection{RL Agent Design}
At the core of the Management layer operates the PPO agent. To effectively train it, the Markov Decision Process (MDP) must be precisely defined through the observation of the environment, also known as state variables, the space of possible actions that the agent can take, and the reward function, which will point its policy towards the optimization objective.

\subsubsection{State Variables}
The network observation must accurately characterize current performance and congestion levels using slice-differentiated metrics alongside global radio channel conditions. Unlike open-source simulators, utilizing real commercial equipment restricts the available state variables to the specific counters supported by the network. For this reason, general metric profiles are defined, from which specific counters are selected based on the infrastructure used for the practical implementation. To manage the coexistence of V2X and eMBB, the following profiles are proposed:

\begin{itemize}
    \item \textbf{Latency (ms):} Critical for assessing V2X SLA compliance. Since true end-to-end (E2E) latency is rarely observable purely from the RAN perspective, latency is estimated using the RLC buffer delay. This metric captures the MAC scheduling waiting time, which constitutes a major component of the overall E2E delay.  
    \item \textbf{Throughput (Mbps):} Serving as the principal eMBB KPI, it is calculated at the Packet Data Convergence Protocol (PDCP) layer by dividing the total volume of data transmitted by the reporting period duration. 
    \item \textbf{Resource Utilization (\%):} Tracking both the PRBs allocated and those actually consumed by each slice is crucial for evaluating allocation effectiveness, allowing the agent to identify potential over-provisioning or resource scarcity. 
    \item \textbf{Channel Quality:} Aggregated cell-level metrics, such as the Channel Quality Indicator (CQI) and applied modulation schemes, are necessary to adapt the policy to the underlying wireless channel variability.
\end{itemize}

In the evaluated four-slice scenario, these profiles instantiate a 31-feature observation vector. For each slice, five features are collected: mean PRB utilization, configured minimum PRB quota, number of active users, PDCP throughput computed from the data volume reported over the preceding five-minute interval, and mean RLC delay. The remaining 11 cell-level features describe radio conditions through CQI and Modulation and Coding Scheme (MCS) histograms, represented using five and six aggregation groups, respectively.

\subsubsection{Action Space}
At each decision point, the agent selects a minimum PRB quota for every active slice. In the four-slice scenario, it selects one quota for each of eMBB-1, eMBB-2, V2X-1, and V2X-2; together, these quotas cannot exceed the cell's total capacity.

While the combinatorial possibilities for PRB quotas are extremely large, training a model with such an extensive action space is infeasible. To accelerate convergence and prevent the exploration of irrelevant configurations (e.g., assigning 0\% of resources to one or more slices), the use of a discretized action space is proposed. This space features a set of 20 predefined quota combinations that span from V2X-prioritized allocations to eMBB-oriented strategies: V2X-1 quotas range from 20\% to 2\%, and V2X-2 quotas from 20\% to 4\%. The quotas assigned to eMBB-1 and eMBB-2 range from 23\% to 41\% and from 35\% to 62\%, respectively. Among the two eMBB slices, eMBB-2 is generally allocated a larger share because of its higher traffic demand (see Table~\ref{tab:traffic_patterns}).

\subsubsection{Reward Function}

To drive the agent's policy, the reward function follows the simulation-based formulation in \cite{PREPRINT-SIM}, which is extended here through closed-loop validation on commercial 5G equipment. The main idea is to enforce the service hierarchy established at the Business Layer. For this case of study, V2X SLA compliance acts as a priority, while eMBB throughput maximization serves as a secondary target: once V2X latency boundaries are satisfied, all remaining resources should be dedicated to eMBB services. To capture this balance, a two-branch reward function is proposed:

\begin{equation}
    R = 
    \begin{cases} 
        \sum\limits_{i \in V2X} -\frac{lat_{i}}{lat_{SLA}} \hspace{0.5em} \text{if } \exists i \in V2X : lat_{i} > lat_{SLA} \\ \\
        -\sum\limits_{i \in \mathcal{S}} \rho_{i} + \sum\limits_{k \in eMBB} \frac{Tp_{k}}{Tp_{k}^{obj}} \hspace{0.5em} \text{otherwise}
    \end{cases}
\end{equation}

The first branch acts as a latency-prioritized penalty: if any V2X slice exceeds its latency threshold ($lat_{SLA}$), the agent receives a severe penalty proportional to the violation, ignoring eMBB performance.

On the other hand, if all latency SLAs are fulfilled, the second branch activates to promote eMBB maximization through two complementary mechanisms: penalizing slice over-provisioning to encourage accurate quotas, and directly rewarding eMBB throughput. This dual approach ensures that the agent actively repurposes unused capacity for eMBB traffic rather than granting V2X slices superior, unrequested performance.

These mechanisms are modeled using two terms. First, $\rho_i$ penalizes PRB over-provisioning:
\begin{equation}
    \rho_i = \max\left(0, \frac{PRB_{asig}^{i} - PRB_{util}^{i}}{PRB_{asig}^{i}}\right)
\end{equation}

This term encourages the allocated quota ($PRB_{asig}^{i}$) to track the observed PRB utilization ($PRB_{util}^{i}$), preventing unpredictable redistribution of spare resources by the DU's MAC scheduler. Concurrently, the final term rewards the achieved eMBB throughput ($Tp_{k}$) relative to the per-slice offered rate ($Tp_{k}^{obj}$, which takes values of 40 and 160 Mbps as shown in Table~\ref{tab:traffic_patterns}). All terms in the function are scaled by weights to provide flexibility.

\begin{table*}[t]
\centering
\caption{Traffic Patterns and Slice Configurations}
\label{tab:traffic_patterns}
\begin{tabular}{@{}llcccl@{}}
\toprule
\textbf{Slice} & \textbf{Profile} & \textbf{UEs Load 1} & \textbf{UEs Load 2} & \textbf{SLA} & \textbf{Generation Params (Per UE)} \\ \midrule
\textbf{V2X-1} & CAM/DENM & 48 & 42 & 20 ms & Period: $\sim$100ms; Size: 300B (CAM); Uniform 200-2000B (DENM)\\
\textbf{V2X-2} & Platooning & 12 & 18 & 20 ms & Period: 10ms; Size: 800B (80\%) or 1200B (20\%) \\
\textbf{eMBB-1} & CBR (Heavy) & 2 & 2 & Max. Thput. & CBR - 20 Mbps \\
\textbf{eMBB-2} & VBR (Heavy) & 2 & 2 & Max. Thput. & VBR- 80 Mbps \\ \bottomrule
\end{tabular}
\end{table*}

\section{Implementation and Experimental Assessment}

\subsection{Experimental Testbed and Scenario}
To validate the proposed framework under real-world constraints, a real 5G SA testbed was deployed. The infrastructure features a fully virtualized Nokia 5G Core and a RAN operating in the n78 band (3.5 GHz) with a 50 MHz bandwidth and 30 kHz subcarrier spacing. The MAC scheduler in the DU natively supports dynamic PRB quota configurations per slice via specific S-NSSAI tags. Moreover, the Network allows a report period of 5 minutes for the collection of PM counters through API requests. 

To emulate the User Equipment (UE) ecosystem, an Amarisoft UEsimbox is employed, capable of simulating up to 64 concurrent high-fidelity users connected to the cell. Finally, an Edge server acts as both the traffic generator and the execution host for the Slice Management framework, fetching metrics, processing them and configuring PRB partitions based on the actions taken by the RL agent.

The evaluation scenario consists of four slices, two eMBB and two V2X, each characterized by distinct user traffic profiles, as summarized in Table \ref{tab:traffic_patterns}. On the eMBB side, heavy data volumes represent services such as high-definition streaming or file transfers. To emulate this traffic, the \texttt{iperf3} tool is utilized, applying a Constant Bit Rate (CBR) profile to eMBB-1, whereas eMBB-2 experiences a bursty Variable Bit Rate (VBR) profile. This combination of eMBB traffic ensures persistent cell congestion, demanding an intelligent resource management.

Conversely, realistic vehicular traffic patterns were extracted from 3GPP TR 37.885 \cite{3GPP-TR-37-885} for the V2X slices:

\begin{itemize}
    \item \textbf{V2X-1:} Standard Cooperative Awareness Messages (CAMs) and Decentralized Environmental Notification Messages (DENMs) represent a high density of users that generate light, periodic traffic.
    \item \textbf{V2X-2:} Emulates a highly connected group of vehicles, consisting of fewer users but transmitting significantly larger data volumes.
\end{itemize}

Furthermore, to evaluate the adaptability of the framework, two distinct load levels are contemplated. In the initial load configuration, V2X-1 supports 48 users and V2X-2 supports 12. In the second load, the distribution shifts to 42 users for V2X-1 and 18 for V2X-2. Due to its light traffic profile, the pressure on the network of V2X-1 remains relatively stable across both configurations. However, V2X-2 experiences a significant increase in data density during the second load level, forcing the RL agent to dynamically adapt its allocation strategy to mitigate the rise in demand. Finally, aligned with the strict requirements of vehicular networks and indications in \cite{3GPP-TR-37-885}, an SLA of 20 ms for both slices was chosen.

\begin{figure}[t]
    \centering
    \includegraphics[width=0.98\columnwidth]{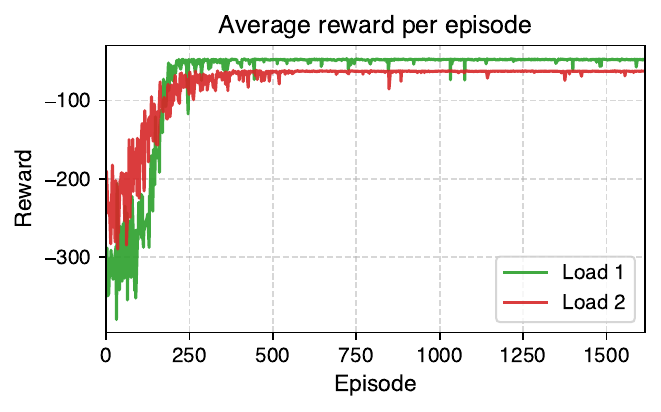}
    \caption{Average reward during training.}
    \label{fig:reward}
\end{figure}

\begin{figure*}[t]
    \centering
    \subfloat[V2X-1 results for baseline cases.\label{fig:baseline1}]{%
        \includegraphics[clip,width=0.98\columnwidth]{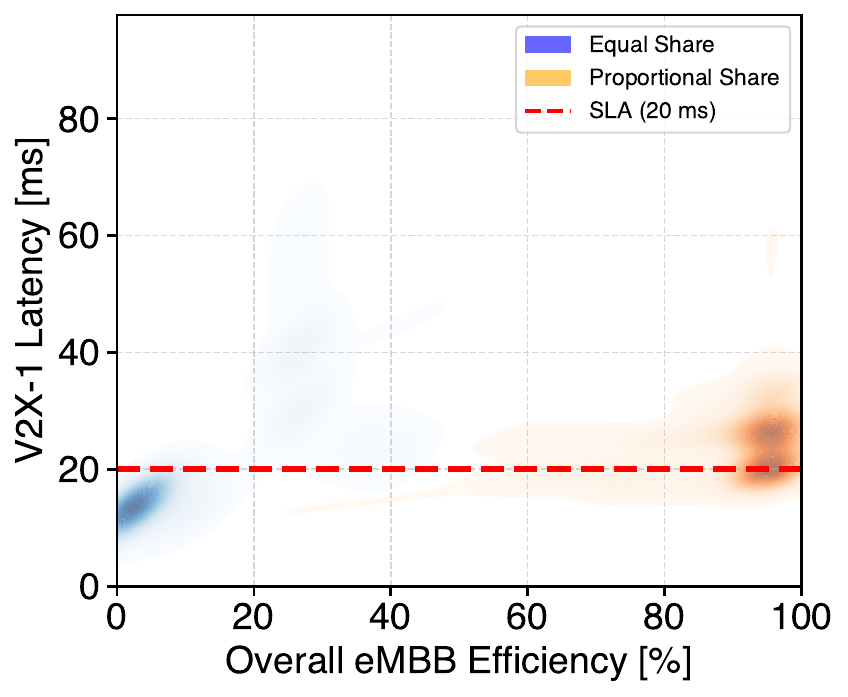}%
    }
    \hfill
    \subfloat[V2X-2 results for baseline cases.\label{fig:baseline2}]{%
        \includegraphics[clip,width=0.98\columnwidth]{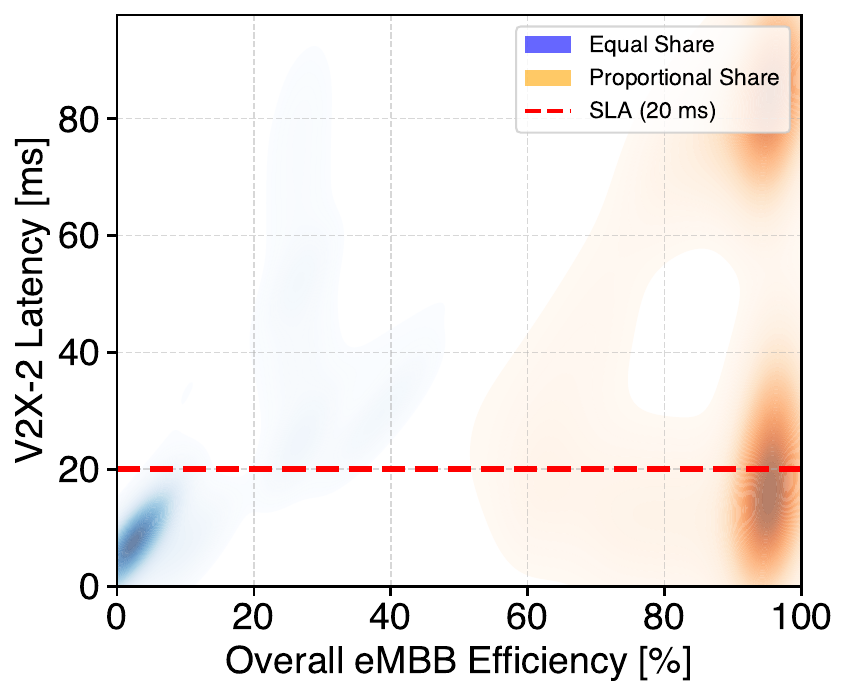}%
    }
    
    \vspace{0.3cm}
    
    \subfloat[V2X-1 results for the proposed DRL-driven system.\label{fig:rl1}]{%
        \includegraphics[clip,width=0.98\columnwidth]{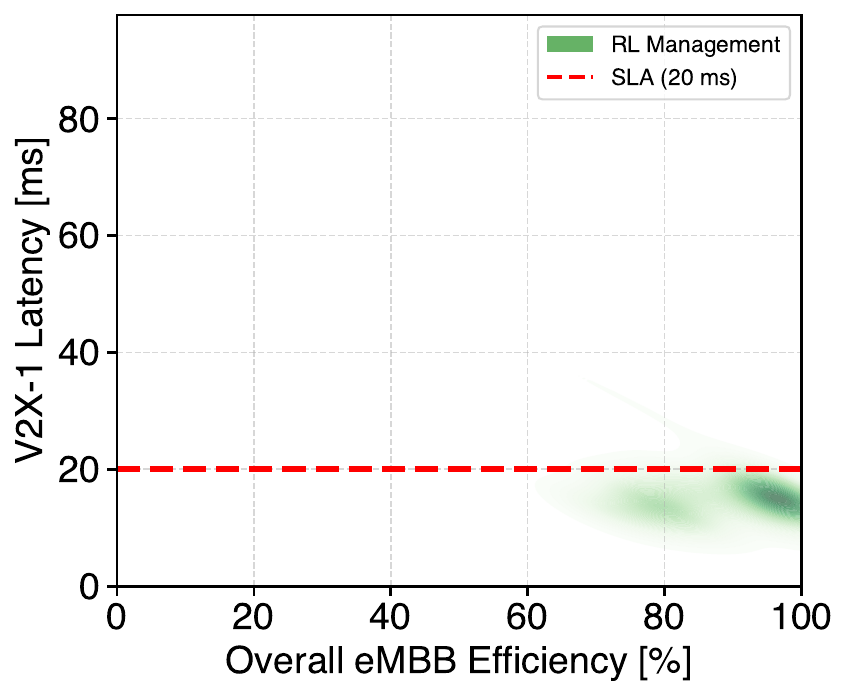}%
    }
    \hfill
    \subfloat[V2X-2 results for the proposed DRL-driven system.\label{fig:rl2}]{%
        \includegraphics[clip,width=0.98\columnwidth]{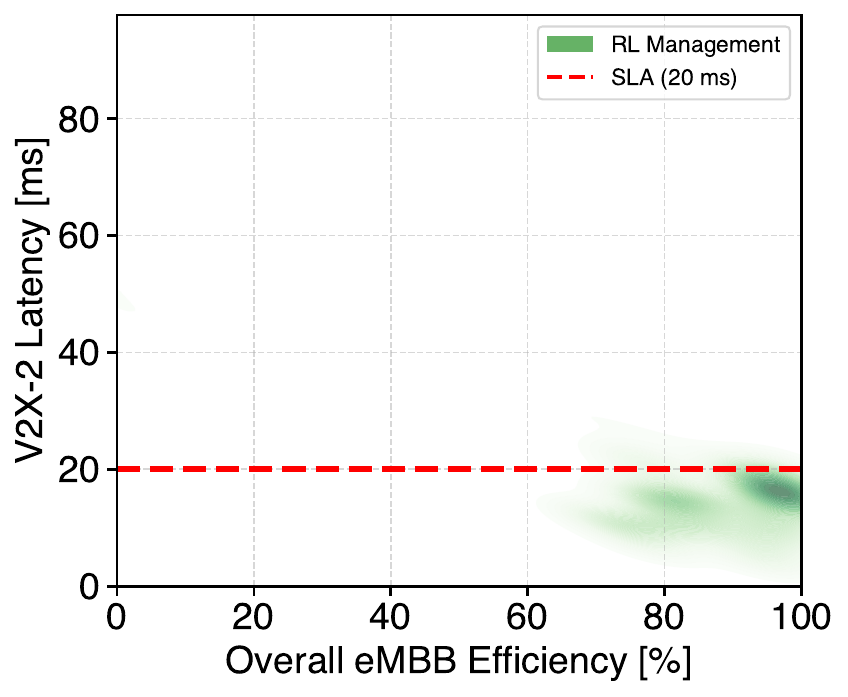}%
    }
    
    \caption{Latency - eMBB Efficiency trade-off comparison for V2X-1 (left) and V2X-2 (right) services. Equal share and Proportional share baselines are shown in the first row, while the proposed DRL-driven system is shown in the second row.}
    \label{fig:baselines_both}
\end{figure*}

\subsection{Training Phase}
Training the PPO agent requires exhaustive action exploration across thousands of iterations. For this reason, it is impractical to perform online training directly within the live network, where each reporting period takes minutes. Moreover, online exploration in live environments is typically infeasible due to the risk of causing severe service disruptions. To address this, an automated measurement campaign was conducted on the testbed to collect empirical telemetry. For each of the 20 discrete actions and under both load configurations, the network was monitored for 3 hours with a 5-minute reporting interval, producing 36 samples per action-load pair, and 1440 samples in total. Within each action-load pair, samples were randomly divided into training and validation subsets using a 70/30 split.

The PPO agent was implemented using Stable-Baselines3, in a measurement-driven environment based on Gymnasium, and with hyperparameter optimization using Optuna. During training, the traffic load (Load 1 or Load 2) alternated unpredictably at the start of each episode, forcing the agent to adapt its policy rather than overfitting to a single traffic profile. 

The average reward evolution during training is depicted in Fig. \ref{fig:reward}. Despite the initial phase of random exploration, the agent's behavior stabilized at around episode 600. The training process concluded with an early stop around episode 1600, instead of executing the originally planned 2000 episodes. 

\subsection{Online Performance Validation}
Once trained, the agent's policy was frozen (exploration disabled) and deployed onto the live 5G testbed for a continuous 6-hour evaluation. The traffic dynamically fluctuated between Load 1 and Load 2 every 30 minutes. To benchmark the performance, two traditional static allocation strategies were also evaluated: \textit{Equal share}, where slices get equal allocations, and \textit{Proportional Share}, where quotas are assigned proportionally with respect to the data volume that each slice handles.

Fig. \ref{fig:baselines_both} shows the joint distribution of V2X latency and overall eMBB efficiency. This metric specifically measures how closely the observed PRB utilization of the two eMBB slices matches their configured minimum quotas. Thus, a score of 100\% indicates that utilization matches the configured quota for both slices; lower scores indicate larger deviations, whether utilization falls below or exceeds the quota.

As observed in Fig. \ref{fig:baseline1} and Fig. \ref{fig:baseline2}, the Equal Share approach provides an excessive, permanent resource share to V2X slices. While this approach achieves substantially lower V2X latencies, it causes severe over-provisioning, positioning the eMBB efficiency below 10\%. Conversely, the Proportional share strategy minimizes V2X resources to boost eMBB throughput, achieving over 72.2\% efficiency for the 10th percentile. However, its lack of adaptability to traffic fluctuations results in frequent and severe SLA violations, especially for V2X-2 (Fig. \ref{fig:baseline2}), where the abrupt increase of users in Load 2 causes two differentiated lobes, one for each load level. 

Finally, the proposed DRL-driven system (Fig \ref{fig:rl1} and Fig \ref{fig:rl2}) achieves a strong compromise between both KPIs. The dynamic agent continuously redistributes resources in response to the active load, achieving 90th-percentile reporting-window mean V2X delays of 17.44 and 19.79 ms, respectively, both below the configured 20 ms threshold. On the other hand, it effectively liberates unused resources to the eMBB slices, achieving a 10th-percentile efficiency of 75.7\%.

\section{Conclusions and Future Work}
The rapid evolution of V2X communications demands intelligent, flexible networks capable of guaranteeing critical SLAs under highly congested, multi-service environments. This paper presented an RL-driven RAN slicing framework utilizing a PPO agent for dynamic PRB quota allocation. Validated on a real-world 5G SA testbed, the proposed system successfully learned an adaptive policy that outperforms static and proportional allocation strategies. Under the evaluated conditions, the framework maintained 90th-percentile V2X latencies below the 20 ms SLA, while achieving high resource efficiency for high-bandwidth eMBB tenants.

Future work will explore the deployment of the agent within the Open RAN architecture, acting as an xApp/rApp in the RAN Intelligent Controller (RIC) to drastically reduce decision latency while increasing portability thanks to open interfaces. Furthermore, the integration of Digital Twins could enhance the offline training phase, safely bridging the gap between simulated exploration and real-world deployment. This RL-driven framework aligns with the broader vision for 6G networks, which anticipates AI-native architectures to sustainably manage the growing complexity and flexibility of advanced Network Slicing.

\bibliography{BIBLIOGRAPHY}
\bibliographystyle{IEEEtran}

\begin{IEEEbiographynophoto}{Daniel E. Garcia-Fernandez} received the B.Sc. in Telecommunications Technologies Engineering, and a M.Sc. degree in Telecommunication Engineering from the University of Malaga in 2025 and 2026, respectively. He is currently pursuing his Ph.D. degree in Telecommunications Engineering at the University of Malaga. His main research interests include AI applications for cellular and Non-Terrestrial Networks (NTN).
\end{IEEEbiographynophoto}

\begin{IEEEbiographynophoto}{Pablo Vera-Soto}
received the B.Sc. degree in Electronics, Robotics and Mechatronics engineering and a M.Sc. degree in Mechatronics engineering from the University of Malaga in 2021 and 2022, respectively and he is currently pursuing his Ph.D. degree in telecommunications engineering at the University of Málaga. His main research focus in mobile communications, avionics networks and cloud robotics.
\end{IEEEbiographynophoto}

\begin{IEEEbiographynophoto}{Sergio Fortes} received the M.Sc. (2010) and Ph.D. (2017) in Telecommunication Engineering from the University of Málaga (UMA), Spain, where he is currently an Associate Professor. He previously worked with major European space agencies (DLR, CNES, ESA) and Avanti Communications. Having led over 30 industry- and publicly funded projects, his research focuses on advanced algorithms and AI applied to cellular and satellite networks, smart cities, cloud robotics, and healthcare.

\end{IEEEbiographynophoto}

\begin{IEEEbiographynophoto}{M. Martínez}
received the B.S degree in Telecommunications Technology
Engineering from the University of Malaga in 2023 and the
Master’s degree in Telecommunication Engineering at the International University of La Rioja. She is currently pursuing her Ph.D. degree in Telecommunications Engineering at the University of Málaga.  Her research interests include the application of AI techniques for the intelligent management of telecommunication networks.
\end{IEEEbiographynophoto}

\begin{IEEEbiographynophoto}{I. de-la-Bandera}
received the MSc and PhD degrees in telecommunications engineering from the University of Málaga. She joined the Communications Engineering Department of the
University of Málaga, Spain, in 2010. Since then, she has participated in many projects, national and international, concerning radio resource management in mobile networks. She has collaborated with major
mobile operators and vendors.
\end{IEEEbiographynophoto}

\begin{IEEEbiographynophoto}{M. L. Luque}
is an AI-RAN Technical Lead within Ericsson’s Cognitive Network Solutions organization, focusing on AI research and prototyping for advanced network solutions. Her background spans standardization, patent development, and network optimization. She holds M.Sc. degrees from the University of Málaga and Aalborg University, graduating with highest honors from both. She is co-inventor of more than 60 patents.
\end{IEEEbiographynophoto}
   
\begin{IEEEbiographynophoto}{A. Mendo}
received the M.S. degree in Telecommunication Engineering with highest honors from the University of Málaga, Spain, in 2004. Since 2004, he has been a Researcher at Optimi Corporation and joined the Ericsson Group in 2010. Having worked in several research projects, he has authored several papers in international conferences and journals and is a co-inventor of more than 30 patents held by Ericsson. His current research interests include self-organizing networks and artificial intelligence.
\end{IEEEbiographynophoto}
   
\begin{IEEEbiographynophoto}{J. Ramiro}
leads the Research and Innovation team within Cognitive Network Solutions in Ericsson’s Cloud and Software Services, driving mid and long-term R\&D in emerging technologies, advanced concepts, and disruptive network optimization. Juan holds a Telecom Engineering degree from the University of Málaga, a PhD in Electrical and Electronic Engineering from Aalborg University, an Executive MBA from San Telmo Business School, and an Executive Degree in Big Data and Analytics from EOI.
\end{IEEEbiographynophoto}

\begin{IEEEbiographynophoto}{R. Barco}
received the M.Sc. and Ph.D. degrees in Telecommunication Engineering from the University of Malaga, Spain, where she is currently a Full Professor. She has worked at Telefonica, Spain, and at the European Space Agency. As a researcher, she specializes in mobile communication networks and smart cities, having published more than 100 scientific papers, filed several patents, and led projects with major companies.
\end{IEEEbiographynophoto}

\vfill

\end{document}